\documentclass[prd,eqsecnum,aps, showpacs]{revtex4}
\begin{document}
\title{Structure formation on the brane: A mimicry}
\author{Supratik Pal \footnote{Electronic address: 
{\em supratik@cts.iitkgp.ernet.in}}}
\affiliation{Department of Physics and Centre for Theoretical Studies\\
Indian Institute of Technology\\ Kharagpur 721 302, India\\ 
and\\
Relativity and Cosmology Research Centre\\ Department of Physics \\ 
Jadavpur University\\  Kolkata 700 032, India}
\vspace{.5in}

\begin{abstract}

We show how braneworld cosmology with bulk matter can explain structure formation.
In this scenario, the nonlocal corrections to the Friedmann equations  supply a
 Weyl fluid that can dominate over matter at late times 
due to the energy exchange between the brane and the bulk.
We demonstrate that the presence of the Weyl fluid radically changes the perturbation
equations, which can take care of the fluctuations required to account for the 
large amount of inhomogeneities observed in the local universe. Further, we show
how this Weyl fluid can mimic dark matter.
We also investigate the bulk geometry  responsible for the scenario. 

\end{abstract}


\maketitle

\section {Introduction}

Observations suggest that
the expanding universe is  homogeneous and isotropic at scales
larger than $150 h^{-1}$ Mpc.
But the large amount of inhomogeneities observed in the local universe 
needs sufficient and convincing explanation.
The usual cosmological models based on the standard
Friedmann equations require that the baryonic matter
fluctuation $\delta > 1$  today, implying $\delta > 10^{-3}$ at the time of recombination.
This is in direct contradiction with cosmic microwave
background observations by over an order of magnitude. This limitation
of standard cosmology leads to an inevitable prediction of dark matter \cite{dm}. 
CMB Observations \cite{cmb} suggest that $\Omega_{\text{baryon}} h^2 \approx 0.02$ {\em i.e.},
only about $4 \%$ of cosmic density is baryonic.
Hence dark matter, if it exists, has to be the dominant nonbaryonic component, 
contributing to as much as $1/3$rd of cosmic density \cite{dmde}. 
It is proposed that its nonbaryonic nature helps it
decouple from radiation, resulting in a growth of structure that starts much before the 
hydrogen recombination. But there are several problems associated with dark
matter, the most pronouncing of which being its ill-response to detection
by series of experiments. Though the experiments of DAMA group \cite{dama+}
have reported in favor of its existence in form of weakly interacting massive particles
(WIMP) \cite{wimp}, similar searches by a number of other groups, such as
 CDMS \cite{cdms}, CRESST \cite{cresst}, EDELWEISS \cite{edel}
 have yielded negative results (for recent results, see \cite{dama-}).
So, there arise questions such as whether dark matter
exists at all or it is the gravity sector,
rather than the matter sector, that needs modifications.
This leads people to consider modified gravity theories, \textit{eg},
modified Newtonian dynamics (MOND) \cite{mond}, bifurcating gravity \cite{bifurc}, 
phantom cosmology \cite{phant} etc.

One such modified gravity theory is the braneworld gravity \cite{rs} 
which has opened up new avenues
of explaining the observations with the help of a modified version of
the standard Einstein equation \cite{eee}. 
In this scenario, when the bulk consists of matter, the bulk metric
for which the FRW geometry on the brane is recovered, is given by a 
higher dimensional generalization of the \textit{radiative} Vaidya
black hole \cite{vaid} that exchanges energy with the brane \cite{maartbulk, lang}.
In presence of bulk matter the \textit{modified} Einstein equation
on the brane reads \cite{maartbulk} 
\begin{equation}
G_{\mu\nu} = - \Lambda g_{\mu\nu} + \kappa^2 T_{\mu\nu} +
6\frac{\kappa^2}{\lambda} {\cal S}_{\mu\nu} - {\cal E}_{\mu\nu} + {\cal F}_{\mu\nu}
\label{eqb9}
\end{equation}
where ${\cal S}_{\mu\nu}$, ${\cal E}_{\mu\nu}$ and ${\cal F}_{\mu\nu}$
are the quadratic contribution from brane energy-momentum tensor,
the projected bulk Weyl tensor and the projected bulk energy-momentum
tensor on the brane respectively.

Incorporating all the braneworld corrections, one can conveniently express the
Friedmann equations on the brane as 
\begin{eqnarray}
H^2 &=& \frac{\kappa^2}{3} \rho^{\text{eff}} + \frac{\Lambda}{3} -
\frac{k}{a^2}  \label{eqb17} \\
\dot H &=&-\frac{\kappa^2}{2} (\rho^{\text{eff}} + p^{\text{eff}}) +
\frac{k}{a^2}
\label{eqb18} 
\end{eqnarray} 
where the \textit{effective} density and pressure are given by \cite{chamb}
\begin{eqnarray}
\rho^{\text{eff}} &=& \rho +\frac{\rho^2}{2\lambda} + \frac{C(t)}{a^4} 
\label{eqb15} \\ 
p^{\text{eff}} &=& p  + \frac{\rho}{2\lambda} (\rho +2p)+\frac{C(t)}{3 a^4}
\label{eqb16}  
\end{eqnarray}
The term $\rho^* = C(t)/ a^4$ is the combined effect of ${\cal E}_{\mu\nu}$
and ${\cal F}_{\mu\nu}$, and is called the ``Weyl fluid", that supplies an
additional perfect fluidlike effect to the usual brane  fluid. Further,
the Weyl parameter $C(t)$ is related to the  black hole mass,
that results in  energy-exchange between the bulk  and the brane 
so that the individual matter-conservation is no longer valid but
the total mass-energy of the bulk-brane system is now conserved. 
Consequently, the matter conservation equation on the brane is no longer a
sacrosanct equation. Rather, it is modified to the  \textit{nonconservation equation} \cite{maart1}
\begin{equation}
\dot\rho + 3 \frac{\dot a}{a} (\rho + p) = -2 \psi
\label{eqb10} 
\end{equation}
where $\psi = d m /d v$, with $m(v)$ the mass of the Vaidya black hole.
It shows how the brane either loses ($\psi > 0$) 
or gains  ($\psi < 0$) energy in exchange with the bulk black hole.
To a braneworld observer, the term $\psi$ is the quantitative
estimate of the brane-projection of the bulk energy density.

The goal of the present article is to 
 show that the bulk-brane energy-exchange results in the growth of the Weyl fluid 
at late times. Following  Newtonian analysis of perturbations from
gravitational instability, which is the simplest yet logical analysis of
gravitational perturbations, we demonstrate that the Weyl  fluid
can mimic dark matter in explaining structure formation. 
That the braneworld gravity can be a very good alternative to dark 
matter in astrophysical contexts of clusters and galaxies 
was proposed in \cite{altdm,  harko}. 
In this article we address the cosmological sector of dark matter.

The plan of the paper is as follows. In Section-II, we  obtain the 
perturbation equations for the effective perfect fluid on the brane.
Section-III is devoted to the solution of the effective perturbation equations
with the help of Weyl fluid 
that mimics dark matter, followed by a comparative analysis with 
the other dark matter models from standard cosmology as well as modified gravity theories.
 We construct the bulk geometry for this setup
 in Section-IV. Finally, we summarize our results and discuss some open issues.


\section{Effective perturbation equations}

Since our focus is on the late time behaviour in the \textit{matter-dominated} era,
we restrict ourselves to the analysis of the zero brane cosmological constant 
scenario. Hence, the equations
of hydrodynamics that involve the quadratic brane correction and the Weyl fluid
correction to the brane perfect fluid, are
\begin{eqnarray}
\frac{\partial \rho^{\text{eff}}}{\partial t} + \overrightarrow\nabla. 
(\rho^{\text{eff}} ~\overrightarrow v^{\text{eff}}) = 0 
\label{eqc1} \\
\frac{\partial \overrightarrow v^{\text{eff}}}{\partial t} + 
(\overrightarrow v^{\text{eff}}.\overrightarrow\nabla)
\overrightarrow v^{\text{eff}} = - \frac{\overrightarrow\nabla 
p^{\text{eff}}}{\rho^{\text{eff}}} -\overrightarrow\nabla \Phi^{\text{eff}} 
\label{eqc2} \\
\nabla^2 \Phi^{\text{eff}} = 4 \pi G \rho^{\text{eff}} 
\label{eqc3} 
\end{eqnarray}
where $\overrightarrow v^{\text{eff}}$ is the velocity
field in the {\em effective} perfect fluid.
It should be noted that the term $\Phi^{\text{eff}}$ is not the usual
Newtonian potential but the effective gravitational potential
which is the resultant effect of the Newtonian as well as
the relativistic potential. The later plays a crucial role  
in the braneworld context and has been discussed in details in \cite{altdm}.
We next consider small perturbations 
\begin{eqnarray}
\rho^{\text{eff}}(\overrightarrow x, ~t) &=& \bar\rho^{\text{eff}}(t) 
(1 + \delta^{\text{eff}}(\overrightarrow x, ~t)) 
\label{eqc4} \\
 \Phi^{\text{eff}}(\overrightarrow x, ~t) &=& \Phi_0^{\text{eff}} + \phi^{\text{eff}}
 \label{eqc5} 
\end{eqnarray}
where $\bar\rho^{\text{eff}}(t)$ and $\Phi_0^{\text{eff}}$ are respectively the
unperturbed effective density and effective  
potential and $\delta^{\text{eff}}$
and $\phi^{\text{eff}}$ are their corresponding fluctuations. 
It is worthwhile to mention the significant difference of the
density fluctuation of the braneworld cosmology
from that of the standard cosmology. In the standard cosmology,
$\delta$ is the fluctuation of baryonic matter only. On contrary, 
in the braneworld cosmology, $\delta^{\text{eff}}$ if the sumtotal
of the fluctuations of baryonic matter and of the  contribution
from braneworld corrections.

We proceed by expressing in terms of comoving coordinates
and neglecting terms of second or higher order, which results
in the following set of simplified perturbation equations 
\begin{eqnarray}
\frac{\partial \delta^{\text{eff}}}{\partial t} + \frac{1}{a}\overrightarrow\nabla_r.
\overrightarrow u^{\text{eff}}= 0 
\label{eqc6} \\
\frac{\partial \overrightarrow u^{\text{eff}}}{\partial t} + \frac{\dot a}{a} 
\overrightarrow u^{\text{eff}}
= - \frac{1}{a} \frac{\overrightarrow\nabla_r p^{\text{eff}}}{\bar \rho^{\text{eff}}}
- \frac{1}{a} \overrightarrow\nabla_r \phi^{\text{eff}} 
\label{eqc7} \\
\nabla_r^2 \phi^{\text{eff}} = 4 \pi G a^2 \bar\rho^{\text{eff}} \delta^{\text{eff}}
\label{eqc8} 
\end{eqnarray}
and express the solution in terms of Fourier transform
\begin{eqnarray}
\delta^{\text{eff}}(\overrightarrow x, ~t) = \sum \delta_k^{\text{eff}}(t) 
~e^{i \overrightarrow k. \overrightarrow x} 
\label{eqc9} \\
\delta_k^{\text{eff}}(t) = \frac{1}{V} \int \delta^{\text{eff}}
(\overrightarrow x, ~t) ~e^{- i \overrightarrow k. \overrightarrow x}
~d^3 \overrightarrow x
\label{eqc10} 
\end{eqnarray}

Further, assuming the effective pressure to be a function of the
effective density alone, the equations of hydrodynamics now
transform into a linear perturbation equation
for $\delta_k^{\text{eff}}$ :
\begin{equation}
\frac{d^2 \delta_k^{\text{eff}}}{d t^2} + 2\frac{\dot a}{a}
\frac{d \delta_k^{\text{eff}}}{d t} - \left(4 \pi G \bar\rho^{\text{eff}}
- (c^2_{\text{eff}} k /a)^2\right)\delta_k^{\text{eff}} = 0
\label{eqc11} 
\end{equation}
where $c^2_{\text{eff}} = \frac{\dot p^{\text{eff}}}{\dot\rho^{\text{eff}}}
= \left[c_s^2 + \frac{\rho + p}{\rho + \lambda} 
+ \frac{4 \rho^*}{9 (\rho + p)(1 + \rho/\lambda)}\right]
\left[1 + \frac{4 \rho^*}{3 (\rho + p)(1 + \rho/\lambda)} \right]^{-1} $ is the
effective sound speed squared \cite{maart1}.


\section{Solutions with the Weyl fluid}

The above perturbative analysis 
can account for the required amount of gravitational instability if the Weyl density 
 redshifts more slowly than baryonic matter density, so that even if
 it starts from a small initial value, it can eventually
dominate over  matter. Now, the nature and evolution of the Weyl fluid is governed
by Eq (\ref{eqb10}) via the 4D Bianchi identity $\nabla^\mu G_{\mu \nu} = 0$. This gives
\begin{equation}
\dot\rho^* + 4 \frac{\dot a}{a} \rho^* = {\cal{Q}}
\label{eqc14} 
\end{equation}
where ${\cal{Q}}$ is a coupling term which can be calculated, 
on principle, if the projected bulk energy density $\psi$ is known. 
But in practice, since no one can fix the exact bulk geometry \textit{a priori},
one has  to take an ansatz for ${\cal{Q}}$ so far as it is physically reasonable
and consistent with the brane equations. One such ansatz has been considered in
\cite{maeda} for a dilaton field in the bulk.  
Let us here take an ansatz  ${\cal{Q}} = \alpha H \rho^*$
($\alpha > 0$), for which the Weyl fluid behaves like
\begin{equation}
\rho^* \propto \frac{1}{a^{(4 - \alpha)}}
\label{eqc16} 
\end{equation} 
 Thus the Weyl parameter is given by  $C(t) = C_0 ~a^{\alpha}(t)$,
 where $C_0$ is its initial value at the matter-dominated epoch.
Obviously, the Weyl fluid is strictly radiationlike only if
$\alpha = 0$, {\em{i.e.}} for matter-free bulk scenario.
But for the bulk with matter, the nature of the Weyl fluid depends on
the coupling strength $\alpha$.
The more the coupling strength $\alpha$ (within the range $1 < \alpha < 4$), 
the more the dominance of the Weyl fluid over matter. 
Hence the Weyl fluid can mimic dark matter for  $\alpha \approx 1$.
However, calculations of CMB anisotropies from the present model
may lead to a better quantitative estimation for $\alpha$.
From Eq (\ref{eqb10}), $\alpha > 0 \Rightarrow \psi > 0$ reveals that the brane 
loses energy to the bulk black hole.
The increase of black hole mass is felt by a braneworld observer through
the projected bulk energy density. 
Consequently,   it results in the growth of the Weyl density
at the expense of brane energy.

We are now in a position of dealing with the perturbation
equation (\ref{eqc11}). This involves the fluctuation of $\rho^{\text{eff}}$ which is a sumtotal
of three quantities given by Eq (\ref{eqb15}). Of them,
the quadratic correction term $\rho^2/ 2\lambda$ 
comes into play at the physics of early universe such as during inflation
(where $\rho \gg \lambda$)  \cite{infl} but it contributes
very little at the present era since $\rho \ll \lambda 
> (100 GeV)^4$. Hence, for all practical purpose, the effective density 
at late times can be approximated as
\begin{equation}
\rho^{\text{eff}} \approx \rho + \rho^* 
\label{eqc12} 
\end{equation}

What turns out from the above equation is that along with the usual matter
density, here we have an additional (Weyl) density contributing to the
total density that governs the perturbation equation  (\ref{eqc11}).
Separating the baryonic (matter) part from the nonbaryonic (Weyl) part of Eq (\ref{eqc11}) 
now yields two wave equations 
\begin{eqnarray}
\frac{d^2 \delta_B}{d t^2} + 2\frac{\dot a}{a} \frac{d \delta_B}{d t} 
= 4 \pi G \bar\rho_B \delta_B + 4 \pi G \bar\rho^* \delta^*
\label{eqc18} \\
\frac{d^2 \delta^*}{d t^2} + 2\frac{\dot a}{a} \frac{d \delta^*}{d t} 
= 4 \pi G \bar\rho^* \delta^* + 4 \pi G \bar\rho_B \delta_B
\label{eqc19} 
\end{eqnarray}
where $\delta_B$ and $\delta^*$ are the fluctuations of baryonic
matter and Weyl fluid respectively
and we have neglected the term involving sound speed because
of the growing fluctuations.  
Consequently, with $\Omega_b \ll \Omega^*$, the relevant growing mode solution for Eq (\ref{eqc19}), 
as a function of the redshift, turns out to be
\begin{equation}
\delta^*(z) = \delta^*(0) (1 + z)^{-1}
\label{eqc20} 
\end{equation}
Substituting the above expression in the fluctuation equation (\ref{eqc18})
of baryonic density  gives 
\begin{equation}
\frac{d^2 \delta_B}{d t^2} + 2\frac{\dot a}{a}
\frac{d \delta_B}{d t} = 4 \pi G \bar\rho^* \delta^*(0) (1 + z)^{-1}
\label{eqc21} 
\end{equation}

Further, we know that though the universe evolves differently at early times,
the standard cosmological solution for the 
scale factor  is recovered in RS-II type brane-world gravity at late times
\cite{maartbulk, maart6}.
So, the late time behaviour for a spatially flat brane is given by
\begin{equation}
a(t) = \left(\frac{3}{2} H_0 t\right)^{2/3(w + 1)} 
\label{eqc13} 
\end{equation}

With this scale factor  and
considering $ \Omega^* \approx 1$ at present time, 
Eq (\ref{eqc21}) now takes the form
\begin{equation}
a^{3/2} \frac{d}{d a} \left( a^{-1/2} \frac{d \delta_B}{d a}\right) 
+ 2 \frac{d \delta_B}{d a} = \frac{3}{2} \delta^*(0)
\label{eqc22} 
\end{equation}

A typical solution for the above equation is given by
\begin{equation}
\delta_B(z) = \delta^*(z) \left(1 - \frac{1+z}{1+z_N}\right) 
\label{eqc23} 
\end{equation}
where we have used the standard relation of the scale factor
with the redshift function $a \propto (1+z)^{-1}$.
Note that though the above relation is a look-alike of the standard
cosmological relation, physically it is completely different. Unlike
the usual dark matter fluctuation, 
here $\delta^*$ is the fluctuation of Weyl density that arises naturally
in  braneworld context.

Let us now analyze  some of the basic features of the present model. 
Eq (\ref{eqc23}) reveals that at $z \rightarrow z_N$, 
the baryonic fluctuation $\delta_B \rightarrow 0$ while
$\delta^*$ remains finite. This implies that even if the baryonic fluctuation 
is very small at a redshift of $z_N \approx 1000$,
as confirmed by CMB data \cite{cmb}, the fluctuations of the Weyl fluid 
had a finite amplitude during that time.  At $z \ll z_N$
the baryonic matter fluctuations are of equal amplitude as
the Weyl fluid fluctuations. This explains the structures we see today.
Further, in this perturbative analysis, no extra matter ({\em {e.g.}} dark matter)
has to be put by hand in order to  explain the structures we see today.

The \textit{effective} equation of state parameter is given by
\begin{equation}
w^{\text{eff}} = \frac{p^{\text{eff}}}{ \rho^{\text{eff}}} 
= \frac{p  + \rho  (\rho +2p)/ 2\lambda + C(t)/ 3 a^4}
{\rho +\rho^2/ 2\lambda + C(t)/ a^4}
\end{equation}
which, in the matter-dominated era,  can be approximated as
\begin{equation}
w^{\text{eff}} \approx \frac{1}{3 (1+ C a^{1 - \alpha})}
\label{eos}
\end{equation}
Clearly, it bears significant difference from the equation of state of cold dark 
matter ($w = 0$). In this sense the term `mimicry' may sound a bit misleading, 
though here we mean that it is the perturbative nature of evolution
that is being mimicked by the Weyl fluid.
Further, $w^{\text{eff}}$ does not cross the phantom divider
line ($w < -1$) \cite{pdl} at least during the matter-dominated era we are
interested about. Hence this model is favored by SNLS \cite{snls}. 
Moreover, the braneworld scenario provides us with a new window for the 
cosmic coincidence problem. Here the evolution of the universe 
may not be a cosmological constant (or dark energy) effect
at all, rather an outcome of the leakage of gravitational signal into 
the extra dimension. Hence the focus now shifts from the coincidence problem
to the question: how can the late-acceleration of the universe be explained
by the modified Friedmann equations ?

A comparative study of our model with the other dark matter models  and modified
gravity theories is instructive.
Eq (\ref{eos}) reveals that our model is distinct from the widely popular
$\Lambda$CDM model \cite{wimp}   so far as the equation of state is concerned. However, 
quite surprisingly, the scale factor bears strong
similarity with that of the matter-dominated phase of $\Lambda$CDM.
Hence, though they have some common features, whether or not the present model will
eventually evolve into $\Lambda$CDM cosmology
at    very late times needs further investigation. As of now, this is an open question
as obvious from the preceeding discussions.
The alternative gravity theory modified Newtonian dynamics (MOND) 
\cite{mond}, is based on modifying Newton's law. Though
Bekenstein has recently proposed a relativistic MOND \cite{mondrel}, a more convincing and
satisfactory relativistic version  bearing important features such as lensing
is yet to come. In comparison, the braneworld model is based on purely relativistic idea and hence 
is potentially more advanced.
Another alternative cosmology is the bifurcating theory \cite{bifurc}
where the effective Lagrangian bifurcates into several  branches, thereby giving
a possibility of unifying dark matter and dark energy. However, here a scalar field
is still required to model dark matter.
The model discussed in the present article do not need any such scalar field. Here the bulk-brane
geometry plays the trick. 
Phantom cosmology \cite{phant} can also provide an unified description
 but it has a wrong sign in the kinetic term that needs convincing physical explanation.


\section{Bulk geometry}

When the bulk consists of matter,  the most general bulk metric for which
a cosmological (FRW) metric on the brane is recovered, is given
by a radiative black hole which is a 5-dimensional generalization 
\cite{maart1, lang,  maartbulk}
of Vaidya black hole \cite{vaid}. In terms of transformed (null)
coordinate $v = t + \int d r/f$, the bulk metric can be written as
\begin{equation}
d S_5^2 = - f(r, ~v) ~dv^2 + 2 dr ~dv + r^2 d \Sigma_3^2 
\label{eqd1} 
\end{equation}
where $\Sigma_3$ is the 3-space. For a spatially flat brane, the function $f(r, ~v)$ is given by
\begin{equation}
f(r, ~v) =  \frac{r^2}{l^2} - \frac{m(v)}{r^2}
\label{eqd2} 
\end{equation}
with the length scale $l$ related to the bulk (negative) cosmological constant
by $\Lambda_5 = -6 / l^2$ and
$m(v)$ is the variable mass of the Vaidya black hole.
This bulk black hole metric is a solution of the 5-dimensional Einstein equation 
 with the bulk energy-momentum tensor 
\begin{equation}
T_{MN}^{\text{bulk}} = \psi q_M q_N
\label{eqd3} 
\end{equation}
where $q_M$ are the ingoing null vectors and $\psi = d m/d v$ is the rate of incoming
radial energy flow to the black hole.

Now, the black hole mass is the rescaled Weyl parameter which is further related to the scale factor by
Eq (\ref{eqc16}) 
\begin{equation}
m(t) = \frac{\kappa^2}{3} C(t) \propto a^{\alpha}(t)
\label{eqd5}
\end{equation}
Hence, with the scale factor of Eq (\ref{eqc13}), the on-brane mass of the bulk black hole 
turns out to be
\begin{equation}
m(t) = m_0 t^{2 \alpha /3(w + 1)} 
\label{eqd6}
\end{equation}
where $m_0$ is the black hole mass at the onset of matter-dominated era,
which is given by
\begin{equation}
m_0 = C_0 \frac{\kappa^2}{3} \left(\frac{3}{2} H_0 \right)^{2 \alpha /3(w + 1)}
\label{eqd7} 
\end{equation}

Since $t$ is the proper time on the brane,
Eq (\ref{eqd6}) gives the on-brane mass $m(t)$. 
In order to obtain the black hole  geometry from the point of interest
of a braneworld observer, 
we have to find out the  off-brane mass $m(v)$. 
However, an exact expression
for the bulk geometry can never be obtained purely from the 
brane data but an approximate expression
for the same at the vicinity of the brane can be obtained by following a perturbative
brane-based approach.

At low energy, the Friedman brane, located outside the ``event horizon",
 moves radially in the bulk. Its radial trajectory
is given by the geodesic $r(t)$, which reduces to the scale factor
$a(t)$ at the brane-location. To a braneworld observer, a brane moving radially 
outwards in the bulk is  identical to an expanding brane  \cite{graviton}.
Hence, the function $f(r,~v)$ at the brane-location reduces to 
\begin{equation}
f(r, ~v)|_{\text{brane}} = \frac{r^2}{l^2} -  C_0 \frac{r^{\alpha}(t)|_{\text{brane}}}{r^2}
= \frac{r^2}{l^2} -  C_0 \frac{a^{\alpha}(t)}{r^2}
\label{eqd7a} 
\end{equation}
where $r(t)|_{\text{brane}}$ is the radial trajectory at the brane-location.
Thus, the null coordinate $v$ turns out to be
\begin{equation}
v = t + \frac{l}{2 \sqrt {m_0} t^{\alpha /3(w + 1)}} \left[\frac{1}{2} 
\ln \left(\frac{r/l - \sqrt {m_0} t^{\alpha /3(w + 1)}}{r/l +
\sqrt {m_0} t^{\alpha/3(w + 1)}}\right) 
+ \tan^{-1} \frac{r/l}{\sqrt {m_0} t^{\alpha/3(w + 1)}} \right] 
\label{eqd8} 
\end{equation}

The off-brane mass $m(v)$ at the vicinity of the brane 
can be found out by expanding $m(v)$ in
Taylor series around its on-brane value $m(t)$ as
\begin{eqnarray}
m(v) =  m(t) + \left[\frac{\partial m}{\partial t_1}\right]_{t_1 = t} \int \frac{d r}{f}
+\frac{1}{2} \left[\frac{\partial^2 m}{\partial t_1^2}\right]_{t_1 = t}
\left(\int \frac{d r}{f} \right)^2 + .....
\label{eqd9} 
\end{eqnarray}

Hence in the present scenario, $m(v)$ can be well approximated as
\begin{equation}
m(v) \approx m_0 t^{2 \alpha /3(w + 1)} + \frac{l \alpha \sqrt {m_0}}{3(w + 1)}
t^{-1 +\alpha /3(w + 1)}
\left[\frac{1}{2} \ln \left(\frac{r/l - \sqrt {m_0} t^{\alpha /3(w + 1)}}{r/l +
\sqrt {m_0} t^{\alpha/3(w + 1)}}\right) 
+  \tan^{-1} \frac{r/l}{\sqrt {m_0} t^{\alpha/3(w + 1)}} \right] 
\label{eqd10} 
\end{equation}

One can now substitute the above expression for $m(v)$ into Eq (\ref{eqd2})
to get the function $f(r,~v)$ near the brane. It is straightforward to construct the
relevant bulk metric at the vicinity of the brane by using together the function 
$f(r,~v)$ and the coordinate $v$ given in Eq (\ref{eqd8}). To avoid the lengthy 
terms involved, we skip the final expression for the metric. 
From the physical point of view, it is not required either. 
The null coordinate $v$ and $m(v)$ given in equations (\ref{eqd8}) and (\ref{eqd10})
respectively will carry all the informations about the bulk geometry to the brane.


\section{Summary and Discussions}

In this article, we have shown that the braneworld cosmology with bulk matter
can explain structure formation. When the bulk is constituted
of matter, then the effective Einstein equation on the brane gives rise to
a quantity that can act as an additional perfect fluid. This so-called
``Weyl fluid'' is the combined effect of the bulk Weyl tensor and bulk energy-momentum
tensor, projected onto the brane. We have shown that the nature of the Weyl fluid 
depends on the energy-exchange between the brane and the bulk so that for
strong bulk-brane coupling, it can dominate over ordinary matter.
Further, following Newtonian analysis of gravitational instability,
we have shown that this Weyl fluid can account for
the required amount of fluctuations in order to explain structure formation. Thus we
conclude that the Weyl fluid can mimic dark matter in structure
formation with the advantage that nowhere we need to introduce any {\em {ad hoc}}
extra matter such as dark matter. We have also investigated the bulk
geometry given by a radiative Vaidya black hole and obtained the
geometric quantities relevant for a braneworld observer.

Throughout this article, we stick to the perturbations from the Newtonian 
gravitational instability. 
Here in no way we tried to study in details the braneworld perturbations as
done for matterfree bulk scenario in numerous papers 
\cite{maart2, maart3, maart4, maart5}.
Rather, we tried to see if the Weyl fluid can mimic dark matter
in cosmological context with the simplest yet logical analysis of
gravitational perturbations. The detailed study of scalar,
metric, curvature, vector and tensor perturbations 
as well as theoretical studies of CMB anisotropies with the Weyl fluid
 behaving as dark matter are left for future works.

An interesting issue is to study gravitational lensing that serves as a
probe of structures. Significant difference in the bending angle due
to the difference in the potentials has been reported in \cite{altdm, harko}. 
It is to be seen how the effective potential affects cosmological lensing.
A comparative study  of the lensing effects with those of the other dark matter
models, {\em {e.g.}} scalar field
haloes \cite{lens1} and verification from \cite{lens2} 
will reveal which properties of dark matter can be reflected by Weyl fluid.  

Finally, we have analyzed structure formation for a flat
universe without considering any accelerated scale factor.
Having realized that braneworld gravity can account
for $\sim 30\%$ of cosmic density usually attributed to
dark matter, one can pay attention to another  $\sim 70\%$  
which is considered to be dark energy
with negative pressure. There currently exists 
some braneworld models of dark energy \cite{de} 
consistent with supernova data. We expect that this perturbative
analysis can be applied to those models, thereby providing an unified
description of dark matter and dark energy from braneworld gravity,
with excitingly new features.


\section*{Acknowledgments} 

I am grateful to Sayan Kar, Narayan Banerjee, Somnath Bharadwaj 
 and Varun Sahni for useful comments.
Special thanks to Ratna Koley for her  encouragements 
and stimulating discussions.

\end{document}